\documentclass[a4paper]{article}

\usepackage{icrc2013}
\usepackage{nth}
\usepackage{textcomp}
\usepackage[english]{babel}

\title{FACT - Threshold prediction for higher duty cycle and improved scheduling}

\shorttitle{FACT - Threshold prediction for higher duty cycle and improved scheduling}

\newcommand{\ethz}{$^1$}
\newcommand{\tudo}{$^2$}
\newcommand{\uniw}{$^3$}
\newcommand{\epfl}{$^4$}
\newcommand{\unige}{$^5$}

\authors{
T.~Bretz\ethz,
A.~Biland\ethz,
J.~Bu\ss\tudo,
D.~Dorner\uniw,
S.~Einecke\tudo,
D.~Eisenacher\uniw,
D.~Hildebrand\ethz,
M.~L.~Knoetig\ethz,
T.~Kr\"ahenb\"uhl\ethz,
W.~Lustermann\ethz,
K.~Mannheim\uniw,
K.~Meier\uniw,
D.~Neise\tudo,
\mbox{A.-K.}~Overkemping\tudo,
A.~Paravac\uniw,
F.~Pauss\ethz,
W.~Rhode\tudo,
M.~Ribordy\epfl,
T.~Steinbring\uniw,
F.~Temme\tudo,
J.~Thaele\tudo,
P.~Vogler\ethz,
R.~Walter\unige,
Q.~Weitzel\ethz,
M.~Z\"anglein\uniw $\;\;$
(FACT Collaboration)
}
\afiliations{
\ethz ETH Zurich, Switzerland $\;\;-\;\;$
   Institute for Particle Physics, Schafmattstr.~20, 8093 Zurich\\
\tudo Technische Universit\"at Dortmund, Germany $\;\;-\;\;$
   Experimental Physics 5, Otto-Hahn-Str.~4, 44221 Dortmund\\
\uniw Universit\"at W\"urzburg, Germany $\;\;-\;\;$
   Institute for Theoretical Physics and Astrophysics,
   Emil-Fischer-Str.~31, 97074 W\"urzburg,\\
\epfl EPF Lausanne, Switzerland  $\;\;-\;\;$
   Laboratory for High Energy Physics, 1015 Lausanne\\
\unige University of Geneva, Switzerland $\;\;-\;\;$
   ISDC Data Center for Astrophysics, Chemin d'Ecogia~16, 1290 Versoix \\
}

\email{thomas.bretz@phys.ethz.ch}

\abstract{
The First G-APD Cherenkov telescope (FACT) is the first telescope using
silicon photon detectors (G-APD aka. SiPM). The use of Silicon devices
promise a higher photon detection efficiency, more robustness and
higher precision than photo-multiplier tubes.  Being  operated during
different light-conditions, the threshold settings of a Cherenkov
telescope have to be adapted to feature the lowest possible threshold
but also an efficient suppression of triggers from night-sky background
photons. Usually this threshold is set either by experience or a
mini-ratescan. Since the measured current through the sensors is
directly correlated with the noise level, the current can be used to
set the best threshold at any time. Due to the correlation between the
physical threshold and the final energy threshold, the current can also
be used as a measure for the energy threshold of any observation. This
presentation introduces a method which uses the properties of the moon
and the source position to predict the currents and the corresponding
energy threshold for every upcoming observation allowing to adapt the
observation schedule accordingly.
}

\keywords{FACT, G-APD, silicon photo sensor, focal plane}

\begin{document}
\maketitle

\section{Introduction}


Since Oct.\ 2011, the FACT Collaboration is operating the
First G-APD Cherenkov Telescope (FACT,~\cite{bib:design}) at the Observatorio
del Roque de los Muchachos at the Canary Island of La Palma.
It is the first focal plane installation based on silicon
photo sensors. The camera, comprising 1440 individually 
read out channels, has been developed to prove the
applicability of Geiger-mode avalanche photo-diodes as
focal plane photo sensors, but also to serve as a monitoring device
at TeV energies for the brightest known blazars. 

While the number of bright sources is limited, the 
number of different environmental conditions during
data taking is infinite. To optimize the distribution
of observation time between the possible sources,
in terms of sampling density, understanding the relation between
the observation conditions and the energy threshold, respectively 
sensitivity, is important.

\section{Concept}

Cherenkov telescopes record nano-second light flashes from
the Cherenkov light emitted by particle cascades in the
atmosphere. Primary gamma-rays induce electromagnetic
showers for which the density of the emitted photons,
in the first order, is directly proportional to the
energy of the primary photon. The total number 
seen by the telescope camera is therefore a direct
measure of the energy as well. Since electromagnetic showers
are highly collimated along the direction of the primary particle,
and therefore have a well defined geometry, also the signal density 
correlated with the primary energy. Therefore, the trigger threshold
of the telescope can also be interpreted as 
an energy threshold.

During observation the zenith angle of the observed
source is changing. With increasing zenith angle, the distance
between the shower core, i.e. the height at which most of the
light is emitted, and ground is increasing.
The light density on ground and with it the light yield 
measured by the telescope is decreasing. This increases the
corresponding energy at the trigger threshold.

The trigger threshold is determined by triggers from random
coincidences of photons from background light, mainly
the diffuse night-sky background and moon light, but is also
influenced by atmospheric conditions like dust in the air or
clouds.

As the ideal trigger threshold, usually the point is assumed
at which the rate from these random triggers becomes negligible
compared to the rate induced by showers from the mainly hadronic
background. 
Due to a very strong dependence of the trigger rate induced
by artificial coincidences of the background photons from the
trigger threshold, this point can be considered independent
from the zenith angle and the highest background spectrum 
close to zenith can be assumed.

If the zenith angle, and therefore the conversion factor 
from light yield to energy is known and the increase of
the trigger threshold can be predicted from the expected
light conditions, the change in energy threshold can also be predicted.

\section{Current prediction}

To be able to predict the trigger threshold, in a first step, 
the flux of background photons must be known. Since this
flux is dominated by moon light, and represented by the
measured current in the camera, a correlation between those
two is searched. Of course, the existence of such a correlation
depends on the stability of the gain of the photo senors.
The stability is discussed in~\cite{bib:stability,bib:moon}.

Empirically, a correlation between the moon-brightness, expressed
as the illuminated fraction of the moon's disc and its zenith angle,
and the measured median current in the camera could be found. 
Further correlations, for example, of the current from
the distance between the object and the moon or the object's
zenith angle are very weak in the dataset used. For the study,
a careful data selection has been done to avoid any bias from
atmospheric conditions, e.g.\ dust, or weather conditions like
clouds. Under real observation conditions, which are not 
always ideal, the current might still be higher than the 
predicted value, but should by no means be lower.

\begin{figure}[t]
 \centering
 \includegraphics[width=0.49\textwidth,angle=0,clip,trim=0 0 0 0]{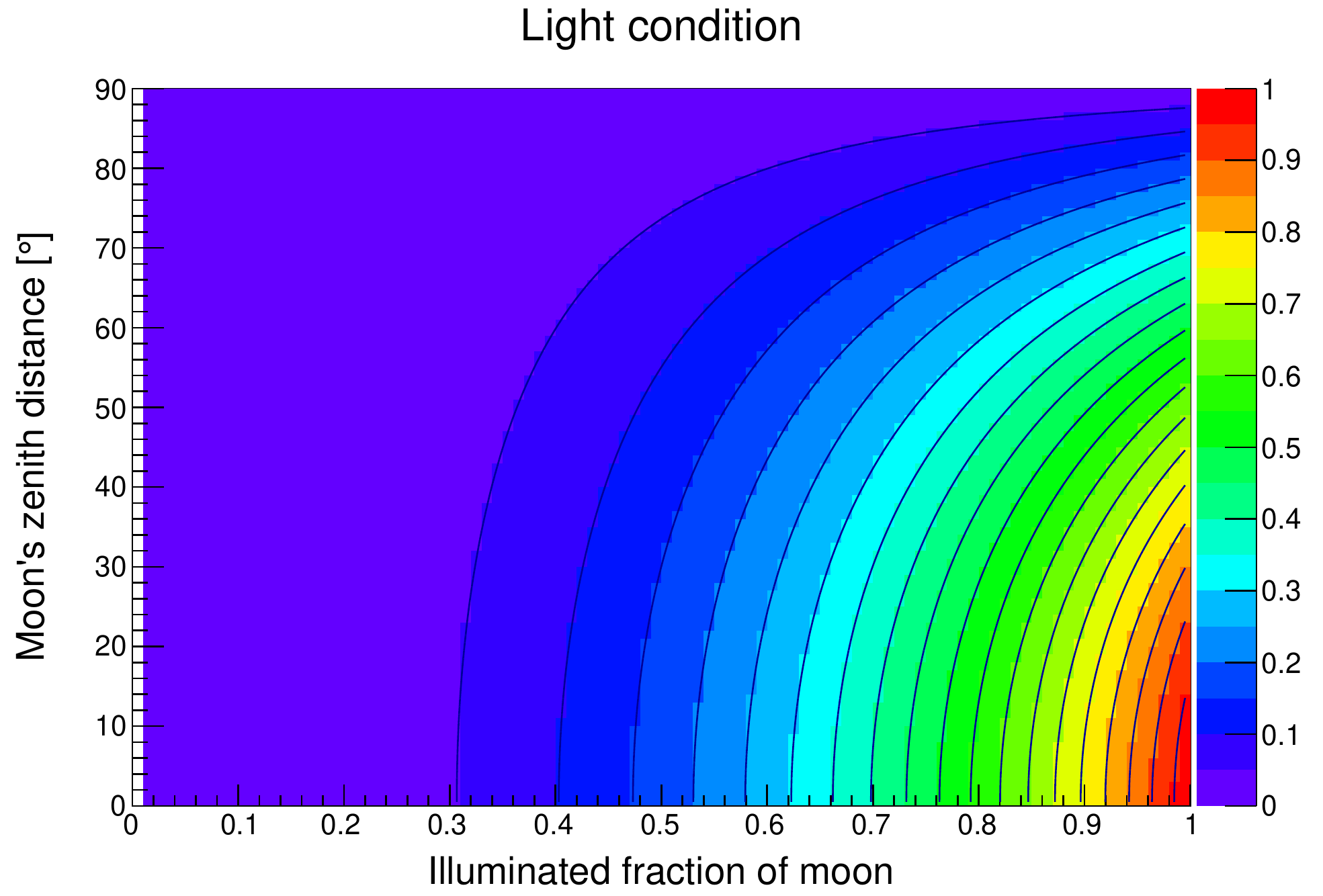}
 \caption{Light condition. As color scale, a value describing the
background light condition is plotted versus the moon's illuminated
fraction and the moon's zenith angle. It is proportional
the measured median current in the camera. Each contour level 
corresponds to a step of 0.05. The values range from new moon
condition (0) to the worst possible condition at full moon (1).}
 \label{fig:lc}
\end{figure}

In Fig.~\ref{fig:lc} a value, called {\em light condition}
directly derived from the moon properties and proportional to the
median current is shown. While a value of zero corresponds
to new moon condition, a value of one would correspond
to full moon with the moon at zenith.

A more detailed discussion can be found in~\cite{bib:moon} and
soon in~\cite{bib:feedback}.

\section{Trigger threshold prediction}

To convert the predicted current into a trigger threshold,
the function describing the threshold dependency of the
rate must be parametrized as a function of the current.
To do this, the dependence of the rate from the threshold
has been measured at different light conditions at ideal
atmospheric conditions and close to zenith. The effect of
atmospheric conditions on ratescans is discussed in~\cite{bib:ratescans}.

\begin{figure}[t]
 \centering
 \includegraphics[width=0.49\textwidth,angle=0,clip,trim=0 0 0.8cm 1.2cm]{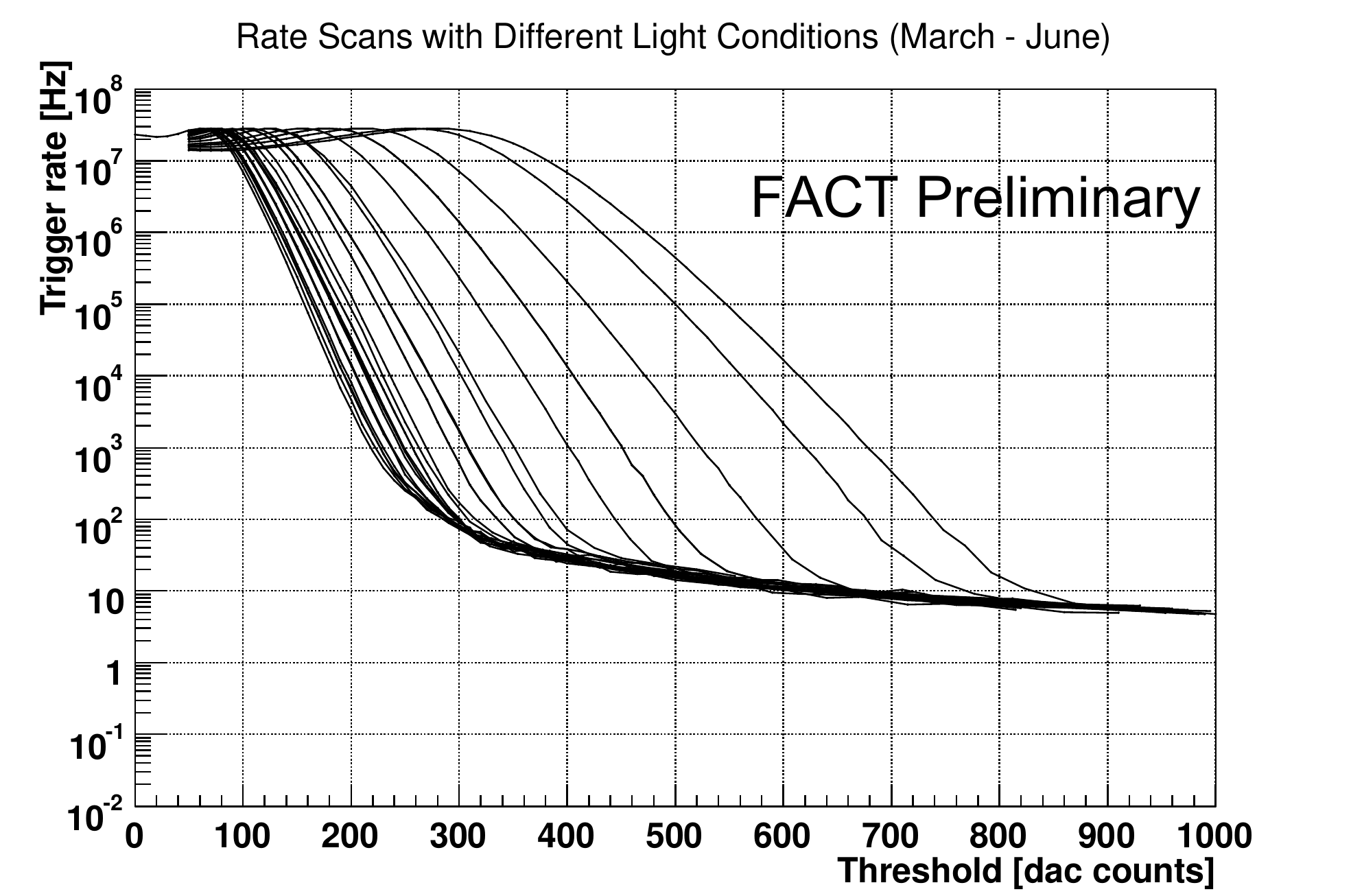}
 \includegraphics[width=0.49\textwidth,angle=0,clip,trim=0 0 0.8cm 0.0cm]{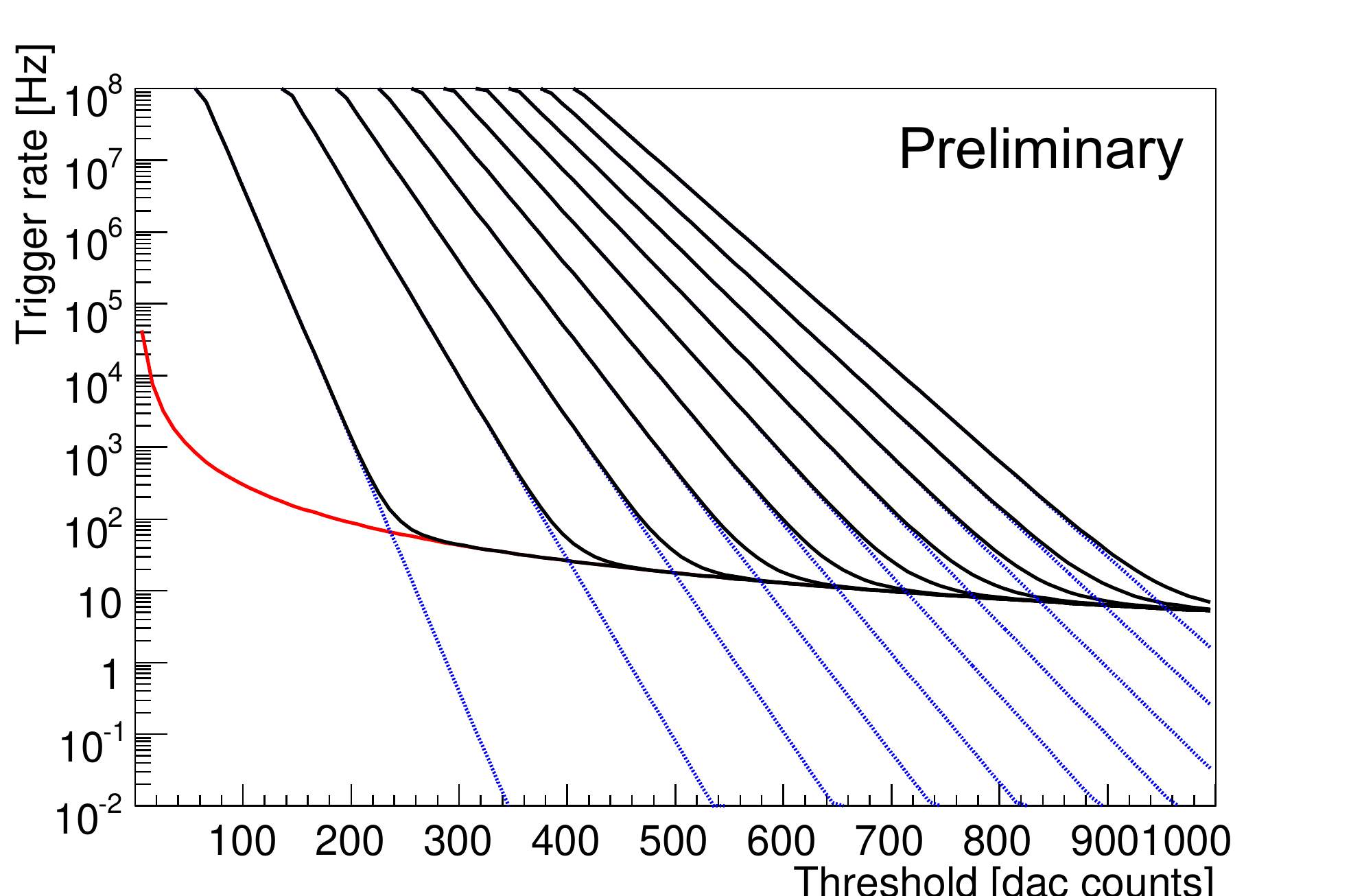}
 \caption{Top: Ratescans taken close to zenith at different light conditions between dark night conditions and almost full moon conditions. Bottom: Parametrization for ratescans depending on the measured median current with 10\(\mu A\)/ch spacing. The power-law of the proton spectrum (red) and the background spectra (black).}
 \label{fig:ratescans}
\end{figure}

In Fig.~\ref{fig:ratescans} (top) some exemplary ratescans are shown.
After the counter got out of saturation, the fast decrease of random
coincidences with raising trigger threshold is visible and
the transition to the spectrum of coincident triggers from background
showers. While the background spectrum is independent of
the background light, the shift of the falling edge between
new moon conditions (left) and almost full moon conditions
(illuminated fraction 90\%, right) is obvious.
The falling edge has been
parametrized by means of linear regression between \(10^3\)Hz and
\(10^4\)Hz. The spectrum of background shower has been parametrized
by a power-law. Knowing the background spectrum is totally
dominated by hadron induces showers and knowing their spectral
index, compatible with the fit result, the power-law index has
been fixed to -1.7 to reduce the number of free parameters.
Fig.~\ref{fig:ratescans} (bottom) shows the result of the
parametrization for a spacing in current of 10\,\(\mu A\)/ch.

From this parametrization, the ideal trigger threshold is derived
as the point at which the random trigger rate has fallen 1/e-th
below the trigger rate from hadron showers. The resulting
current dependent trigger threshold is now in use during
standard data taking since several months. Before each data run,
which usually lasts five minutes, the median current is readout and
the corresponding trigger threshold set. Except the influence
of bright stars in the field-of-view, the threshold is then kept 
constant during these five minutes of data taking, to allow
for easier analysis of the data.

\begin{figure}[t]
 \centering
 \includegraphics[width=0.49\textwidth,angle=0,clip,trim=0 0 0.6cm 1.3cm]{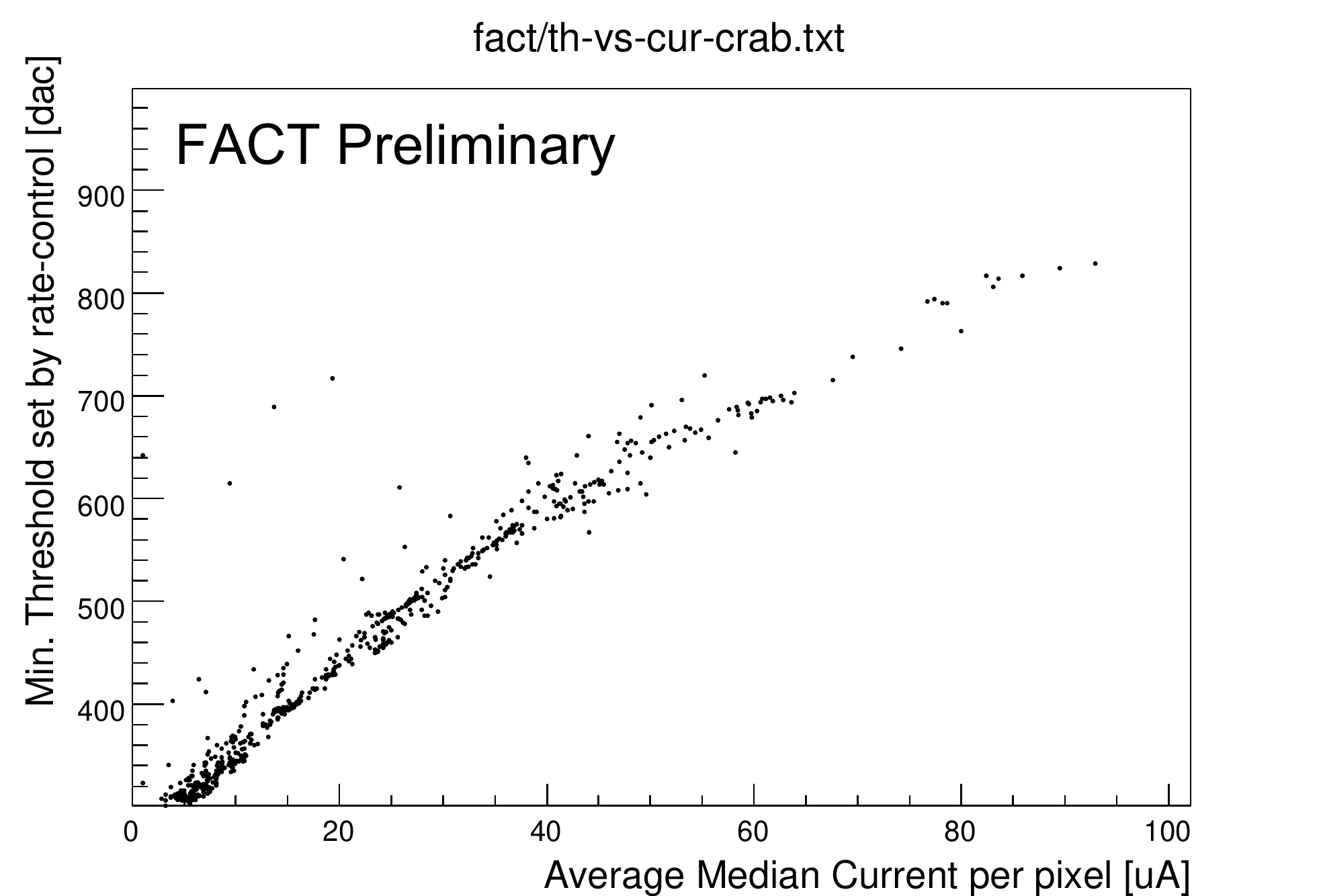}
 \includegraphics[width=0.49\textwidth,angle=0,clip,trim=0 0 0.6cm 0cm]{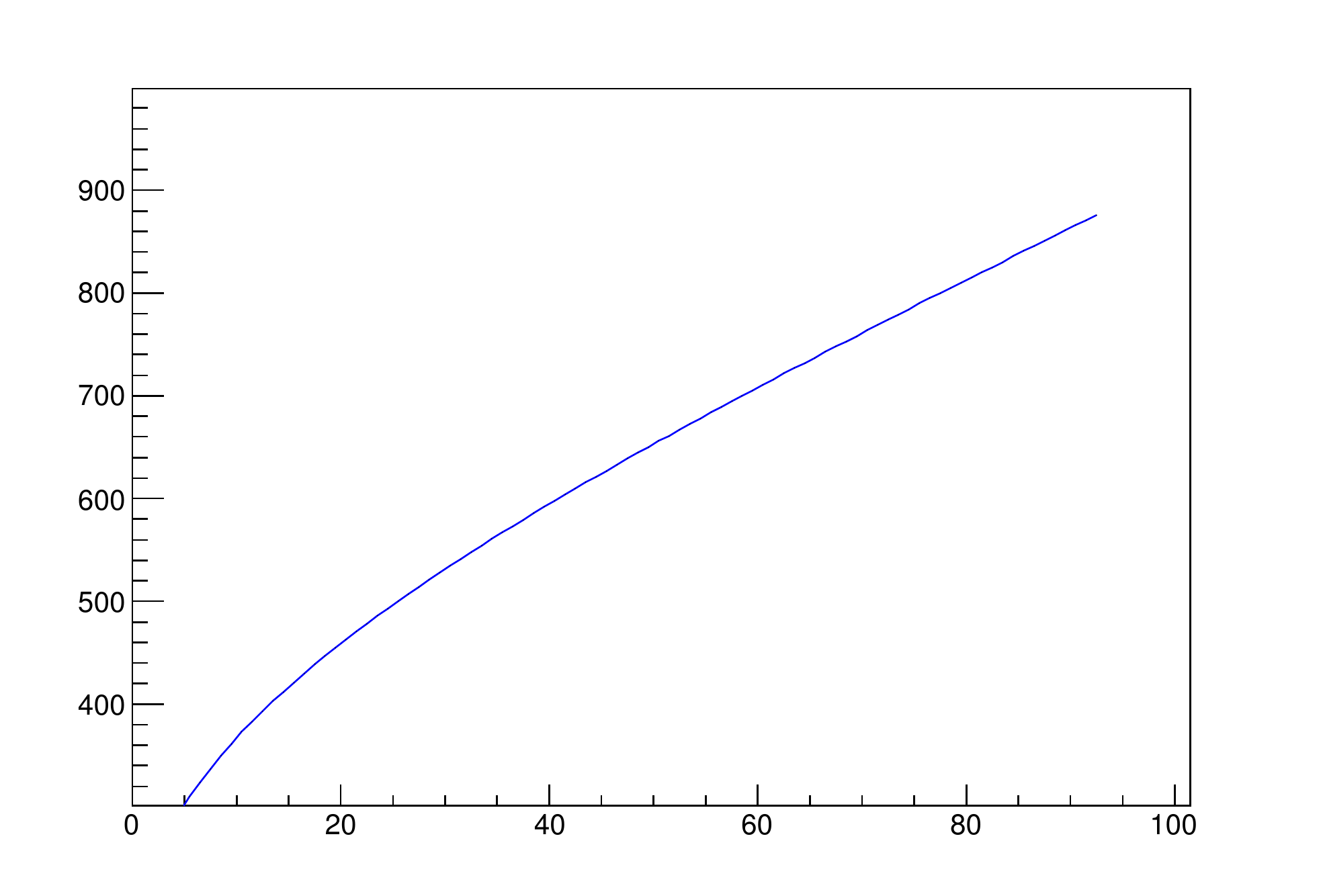}
 \caption{Top: Threshold to archive a trigger rate below 70\,Hz as determined by a mini-ratescan versus median current, applied for several month before each 5\,min data run. Bottom: Trigger threshold depending on median current determined
from the parametrized ratescans shown at the top in Fig.~\ref{fig:ratescans} (bottom) for
a fixed rate of 70\,Hz. A good agreement with the measurement in
 the top figure is visible.}
 \label{fig:measurement}
\end{figure}

Before this method of setting the threshold was implemented,
the trigger threshold during
data taking was determined by a kind of mini ratescan before each run,
which effectively increased the trigger threshold until the rate had 
dropped below 70\,Hz. The obtained correlation between threshold
and median current is shown in Fig.~\ref{fig:measurement} (top). For 
comparison, Fig.~\ref{fig:measurement} (bottom) shows the same
but obtained from the parametrization shown in Fig.~\ref{fig:ratescans} (bottom).

Taking the proportionality between the energy and the
light yield into account, it can be assumed that the
trigger threshold and the energy threshold are proportional as well.
Therefore, the obtained parametrization can be used to derive the
worsening of the energy threshold with increase in current.
The previously obtained correlation, allows to directly correlate
the energy threshold with the moon properties.

\section{Energy threshold prediction}

To adapt the obtained energy threshold to the zenith angle
of the source, the decrease of light density on the ground 
with increasing zenith angle has to be taken into account.

\begin{figure}[t]
 \centering
 \includegraphics[width=0.49\textwidth,angle=0,clip]{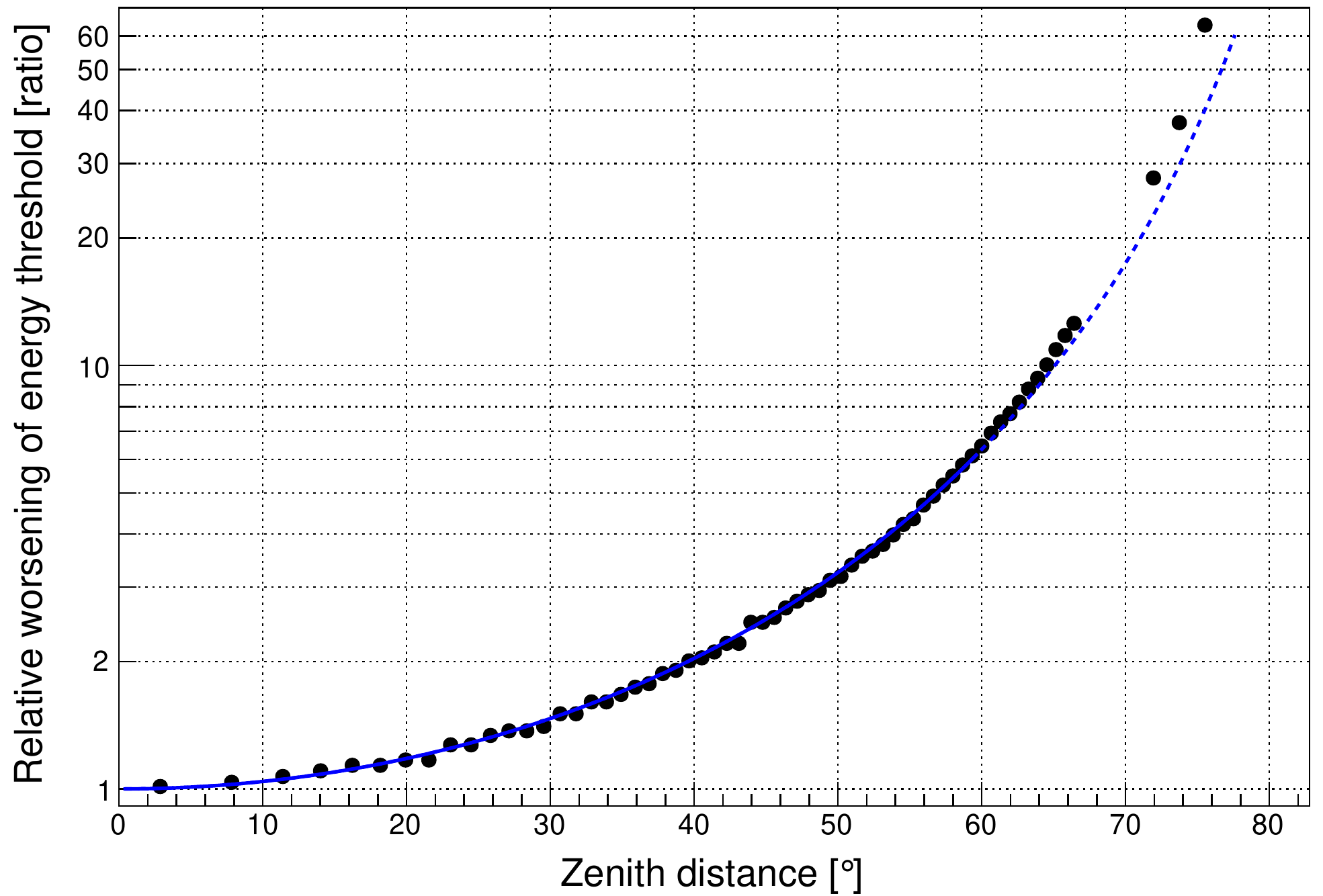}
 \caption{Relative worsening of the energy threshold expressed
as a ratio versus the zenith angle \(\theta\). The black
points are simulation results
for the MAGIC telescope and were taken from~\cite{bib:thesis}.
A fit to the data (up to 60\textdegree{} zenith angle) in \(\cos^\alpha\theta\) is overlayed as
blue line.}
 \label{fig:rel_th}
\end{figure}

This effect can either be calculated or simulated. A Monte Carlo
simulation how the energy threshold changes with zenith
angle at constant trigger threshold can be found in~\cite{bib:thesis}.
The data obtained from this study is shown in Fig.~\ref{fig:rel_th}
as a relative change of energy threshold versus zenith angle.
The data was fitted with \(\cos^\alpha\theta\) yielding 
\(\alpha=-2.66\) which is overlayed as blue line.
Although, geometrically a square dependence is expected,
this result is worsened by the increase of absorption
due to the longer path length at higher zenith angles.
For the fit, only data points up to 60\textdegree{} zenith distance
were taken into account, because starting around this
angle, it makes a significant difference for the result
whether a curved or flat atmosphere is assumed in the simulation.
According to the description of
the simulation, the curved atmosphere was not switched on in the
simulation, which results in an energy threshold at high zenith angles 
worse than expected. However, the difference
up to 70\textdegree{} zenith distance is still at the
few percent level and can be neglected. Observations above
70\textdegree{} would anyway only be scheduled in exceptional cases
in which the precise decrease in energy threshold is not expected
to be the primary criteria anymore.

Although, strictly speaking, this result is not valid for the
FACT telescope because it was obtained for the MAGIC 
telescope, no major difference between both is expected
under the assumption that the total light yield corresponding
to a given trigger threshold does not significantly 
change with zenith angle.

\section{Result}

\begin{figure}[htb]
 \centering
 \includegraphics[width=0.49\textwidth,angle=0,clip,trim=0 0 0.5cm 0]{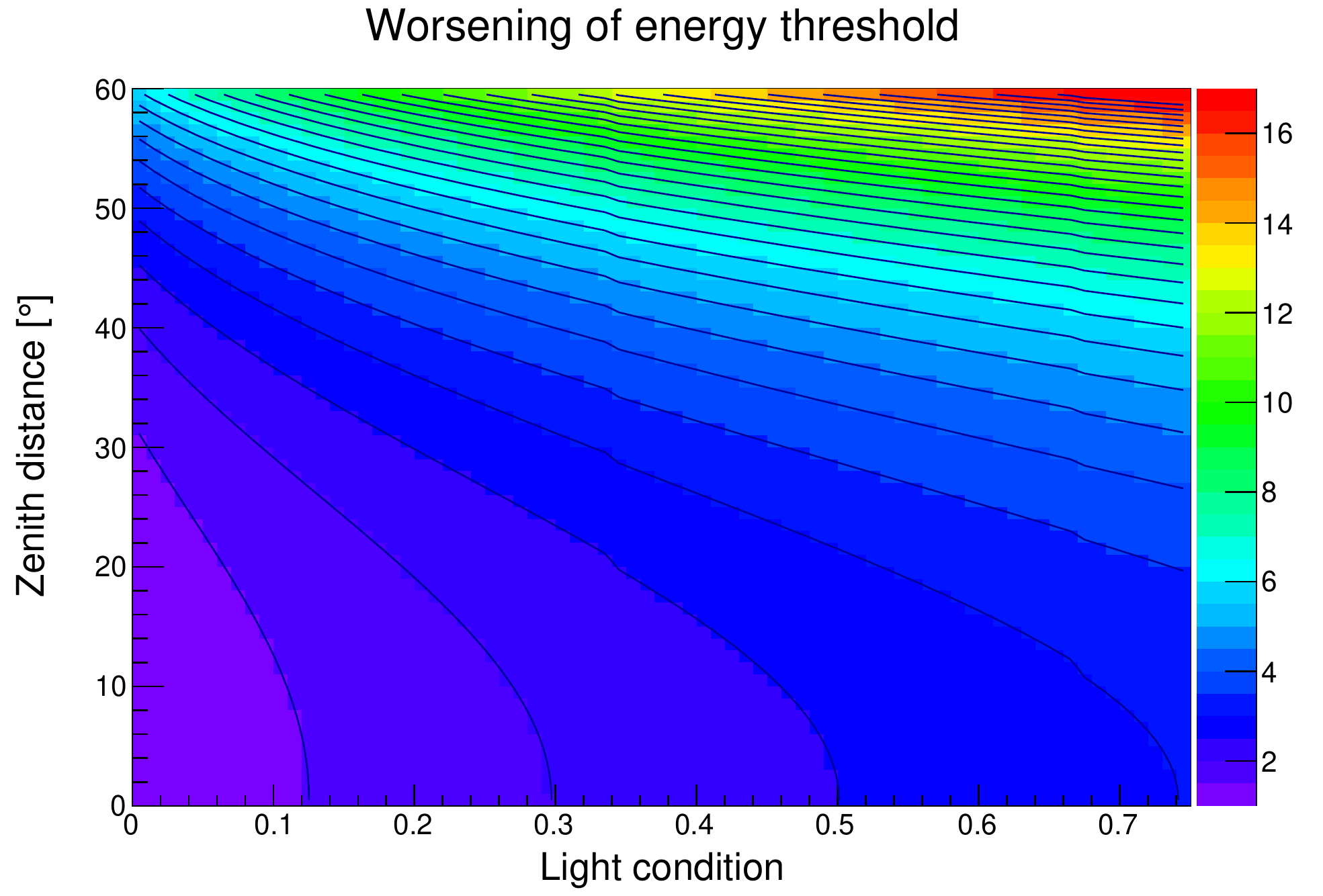}
 \caption{Worsening of energy threshold color coded 
expressed as a ratio
of energies versus light condition and the zenith angle of the
observed source.
Light condition 0 corresponds to a new moon night, light condition
1 would correspond to a full moon night with the moon at zenith.
The full dependency is shown in Fig.~\ref{fig:lc}.
 Each contour level corresponds to a step of 0.5. For example, comparable
energy thresholds are expected for a light condition of 0.75 close to zenith and at 50\textdegree{} zenith angle at dark night. A light condition of 0.75 corresponds
to the moon illuminated by not more than 75\% also close to zenith.}
 \label{fig:eth2d}
\end{figure}

Both results, the increase of the trigger threshold as shown
in~\ref{fig:measurement} depending on the moon's properties, and
the increase of energy threshold depending on the zenith angle of
the source can now be combined into a single prediction for
the worsening in energy threshold. This worsening
is shown in Fig.~\ref{fig:eth2d}. The relative increase in energy,
i.e.\ worsening in energy threshold, is shown as a color scale versus
the light condition and zenith angle of the observed source.
A light condition of 0 corresponds to new moon conditions, while
a light condition of 1 would correspond to full moon with the 
moon at zenith. The full dependency of the light condition on
moon brightness and zenith angle is shown in Fig.~\ref{fig:lc}.
It can be seen that at very bright condition, the brightness
dominates the energy threshold while at high zenith angles,
the background light level has only a minor influence.

\section{Conclusion}

With the FACT telescope, the first complete focal plane
installation using silicon photo sensors is available.
It was proven that even under real observation conditions
like changing weather and light conditions, a stable operation
is possible within the limits of the calibration of the bias
voltage.

Using this stability, it was possible to predict the current
as a representation for the photon flux from the diffuse background
light depending on light conditions, i.e. depending on the
moon's zenith angle and brightness. 

This prediction was then correlated with the measured rate
depending on the trigger threshold setting allowing.
This enables the use of the measured median current in the
camera to define the global trigger threshold during observation
which has turned out to be a very stable and robust procedure.

Given that the threshold is closely correlated to a
local light density in the camera and that gamma showers,
especially at the trigger threshold, are very well concentrated,
the trigger threshold is equivalent to a measured energy.

Putting this result together with the change in energy threshold
coming from the change in light density with different observation
zenith angles, yield a prediction of the worsening of the
energy threshold with light condition and zenith angle.

Since the main physics goal of the FACT telescope is the
long-term monitoring of bright TeV blazars (mainly Mrk\,421 and
Mrk\,501) the prediction of the energy threshold is important
to allow for the optimum balance in sampling density with their
quiescent flux as reference.

Given the stability of the system and the possibility to
optimize the physics output, the FACT telescope does not only
prove the applicability of silicon photo sensor in focal planes and
especially in Cherenkov telescopes, it is also an ideal
instrument to gain knowledge on the long-term behavior
of the brightest blazars.\\

An overview of the telescope hardware and software is given
in~\cite{bib:design}. An detailed discussion of the stability 
will be available soon in~\cite{bib:feedback}.

\vspace*{0.5cm}

\footnotesize{\paragraph{Acknowledgment}{The important contributions
from ETH Zurich grants ETH-10.08-2 and ETH-27.12-1 as well as the
funding by the German BMBF (Verbundforschung Astro- und
Astroteilchenphysik) are gratefully acknowledged. We are thankful for
the very valuable contributions from E.\ Lorenz, D.\ Renker and G.\
Viertel during the early phase of the project We thank the Instituto de
Astrofisica de Canarias allowing us to operate the telescope at the
Observatorio Roque de los Muchachos in La Palma, and the
Max-Planck-Institut f\"ur Physik for providing us with the mount of the
former HEGRA CT\,3 telescope, and the MAGIC collaboration for their
support. We also thank the group of Marinella Tose from the College of
Engineering and Technology at Western Mindanao State University,
Philippines, for providing us with the scheduling web-interface.}}


\begin{thebibliography}{}


\bibitem{bib:design} H.~Anderhub et al. (FACT Collaboration), 
2013, JINST {\bf 8} P06008 [arXiv:1304.1710].

\bibitem{bib:stability} T.~Bretz et al. (FACT Collaboration), these proc., ID\,683.

\bibitem{bib:moon} M.~Koetig et al. (FACT Collaboration), these proc., ID\,695.

\bibitem{bib:feedback} FACT Collaboration, {\em in prep.}

\bibitem{bib:ratescans} D.~Hildebrand et al. (FACT Collaboration), these proc., ID\,709.

\bibitem{bib:thesis} R.~Firpo, thesis, IFAE/UAB 2006.

\bibitem{bib:status} T.~Bretz et al. (FACT Collaboration), these proc., ID\,682.


\end{thebibliography}
\end{document}